# Computational Design of Active 3D-Printed Multi-State Structures for Shape Morphing

**Lumpe, Thomas S.**[1]

Engineering Design and Computing Laboratory

Department of Mechanical and Process Engineering

ETH Zurich

Tannenstrasse 3, 8092 Zurich, Switzerland

tlumpe@ethz.ch

**Tao, Michael**

University of Toronto

Toronto, Canada

mtao@dgp.toronto.edu

**Shea, Kristina**

Engineering Design and Computing Laboratory

Department of Mechanical and Process Engineering

ETH Zurich

Tannenstrasse 3, 8092 Zurich, Switzerland

kshea@ethz.ch

**Levin, David I. W.**

University of Toronto

Toronto, Canada

diwlevin@cs.toronto.edu

---

[1] Corresponding author

**Abstract**

Active structures have the ability to change their shape, properties, and functionality as a response to changing operational conditions, which makes them more versatile than their static counterparts. However, most active structures currently lack the capability to achieve multiple, different target states with a single input actuation or require a tedious material programming step. Furthermore, the systematic design and fabrication of active structures is still a challenge as many structures are designed by hand in a trial and error process and thus are limited by engineers' knowledge and experience. In this work, a computational design and fabrication framework is proposed to generate structures with multiple target states for one input actuation that don't require a separate training step. A material dithering scheme based on multi-material 3D printing is combined with locally applied copper coil heating elements and sequential heating patterns to control the thermo-mechanical properties of the structures and switch between the different deformation modes. A novel topology optimization approach based on power diagrams is used to encode the different target states in the structure while ensuring the fabricability of the structures and the compatibility with the drop-in heating elements. The versatility of the proposed framework is demonstrated for four different example structures from engineering and computer graphics. The numerical and experimental results show that the optimization framework can produce structures that show the desired motion, but experimental accuracy is limited by current fabrication methods. The generality of the proposed method makes it suitable for the development of structures for applications in many different fields from aerospace to robotics to animated fabrication in computer graphics.

# INTRODUCTION

Active materials can enable advanced functionalities such as shape-morphing in structures, but most active structures from literature are limited with respect to the number of achievable target states, usually only one, require different input actuations to achieve multiple states, or require a tedious material programming step (1–4). This makes such methods poorly suited for repeatable transitions between multiple target states. Further, applying a stimulus such as heat is often limited to global activation such as water baths or environmental chambers, which limits the applicability of active structures outside the laboratory environment (5). Instead, local heating can be used to differentially control the material properties in different parts of a structure (6). This allows for mapping multiple state transitions to a single mechanical input actuation, which increases the number of deformations that can be encoded in a single object. However, finding structural configurations and heating patterns that achieve specified target deformations is a complex design task and requires comprehensive computational design methods (7, 8). This specifically includes the need to find structures that are fabricable with current technologies. Adding heating elements such as copper wire coils to a structure after fabrication can provide the necessary flexibility but poses requirements on the minimum size of structural elements as it can be challenging to apply them to intricate lattice topologies. A method that can generate individual cells with minimum dimensions can be found in Computer Science in the form of power diagrams, which are a generalized version of Voronoi diagrams (9). As such, the combination of power diagrams with topology optimization provides the flexibility to form structural patterns that support the integration of multiple target states while explicitly considering the practicability of adding heating elements post-fabrication.

Combining these concepts, this work shows how the temperature sensitive nature of active materials can be exploited to allow an object to deform to different states in response to an identical actuator input. Multi-state transitions are directly programmed into the object itself via both material distribution and temperature control. The multi-state design problem is modeled as a simultaneous material and temperature optimization, which is then solved with a new topology optimization scheme that is based on volume-constrained power diagrams and weighted triangulations. The power-diagram discretization allows for enforcing a minimum size constraint on individual cells to make fabrication and actuation of the optimized structures possible. This work is focused on in-plane, 2D deformations and a number of examples relevant to mechanical engineering and computer graphics are presented. All examples are fabricated by 3D printing of dithered materials using a multi-material 3D printing process to tune the thermo-mechanical properties of the resulting structures. Finally, the fabricated structures are differentially heated by copper coils and mechanically tested to verify the encoded deformation behavior.

# MATERIALS AND METHODS

## Multi-State Structures

This work introduces a new form of multi-state fabrication to encode multiple motions into a single actuated object by exploiting the heat sensitivity of shape memory polymers (SMPs). Popular approaches to 4D printing exploit the special molecular arrangement of SMPs, which allow deformations in the rubbery state to be fixed by cooling the material below the glass transition temperature $T_g$ while holding it in its deformed shape. This is often referred to as the programming step in literature (10). By re-heating the SMP above its $T_g$, the undeformed initial state can be recovered. This is normally achieved via heating in a water bath or with warm air. Recovery to the original state is driven by low magnitude internal forces and repeated actuations or actuations to different states require re-execution of the tedious programming step.

The approach presented in this work significantly differs from the standard 4D printing approach as the motion of the structures is not generated by the recovery forces of the SMP, but by an independent input actuation. Rather than an ambient heat source, copper wires are used to locally control the heating of the structures. This creates local material stiffness gradients, which alter the global deformation mode for a given driving actuation. By selectively heating different parts of the structure, different end states can be generated despite starting from the same initial state of a single printed structure. This approach confers two advantages. First, the structures can transition between states relatively quickly as the transition speed is determined by the speed of the applied driving actuation. And second, structures can repeatedly transition between states without re-programming as the state transitions are encoded once at the moment of fabrication and remain persistent thereafter.

## Design Optimization

The key computational problem in the multi-state fabrication design is one of material placement and heating: how can material be arranged and heated inside the design domain such that objects can achieve the correct deformed shapes? Hence, the input to the optimization framework is a design domain $\Omega$ that describes the initial shape of the multi-state object, a given base material (including mechanical and thermal properties), a set of boundary conditions, a single driving actuation, and a set of target deformations. The output is a material distribution and a heating pattern, which enable the desired target deformations for the prescribed input actuation and boundary conditions. To obtain structures that can be fabricated and actuated, several requirements are imposed on the optimization scheme. As such, the method must:

- Match multiple target states under a given set of boundary conditions.
- Have a binary material distribution.
- Form a single, connected region.
- Limit material feature size to prevent regions that are too small to be printed or actuated.

The overall design approach comprises several steps and is illustrated in the following paragraphs using the example of a "Gingerbread Man" that can lower its left arm or lower its right arm by pushing its head, schematically shown in Fig. 1.

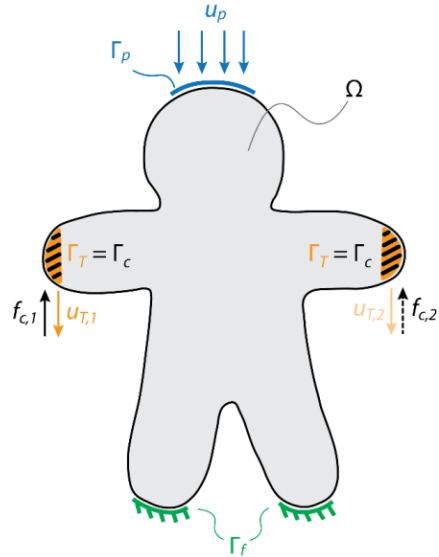

Fig. 1: Optimization problem setup on the example of the "Gingerbread Man" with two target states. The design domain $\Omega$ determines the overall shape of the structure. $\Gamma_f$ indicates the fixed boundary region and $u_p$ denotes the applied mechanical actuation on the domain $\Gamma_p$. The target displacements for the two states are $u_{T,1}$ and $u_{T,2}$ and are specified on the nodes in $\Gamma_T$. Finally, $f_{c,1}$ and $f_{c,2}$ are used to ensure the optimized structure is connected and structurally sound.

**Target Displacement Optimization**

To begin, a standard single-state deformation design problem is considered. The goal is to minimize the pose error $J$, which measures the distance between the displacements $u$ of a body under prescribed boundary conditions and the target displacements $u_T$ evaluated on the target boundary $\Gamma_T$, with the material density distribution $\boldsymbol{\rho}$ of the body as the design variable. The standard $L_2$-norm is used to measure the error between $u$ and $u_T$. This yields the following optimization problem:

$$\min_{\boldsymbol{\rho}} \; J(\boldsymbol{\rho}) = \|u_T - u(\boldsymbol{\rho})\| \tag{1}$$

The structure is fixed at the boundary $\Gamma_f$. This basic optimization problem essentially describes the design of a compliant mechanism as it is often found in literature (11), where an input displacement or force is transferred to an output deformation or force by means of elastic deformation of the structure.

This formulation is easily extended to include multiple deformed states by considering the summation of individual pose errors, $J_j$, one for each set of $k$ target deformations:

$$\min_{\rho} \sum_{j=1}^{k} J_j(\rho) \tag{2}$$

$$\text{with } J_j(\rho) = \|u_{T,j} - u_j(\rho)\|$$

Next, it must be ensured that the deformations of the structure are physically valid. This is achieved by adding a constraint to enforce static equilibrium of the structure. Assuming a linearly elastic constitutive model, this constraint can be discretized using the finite element method (FEM). For a single target state, this yields:

$$\min_{\rho} J(\rho) = \|L_u(u_T - u(\rho))\| \tag{3}$$

$$\text{for } K_\rho u = f$$

where $K_\rho$ is the stiffness operator. The binary diagonal matrix $L_u$ restricts $J$ to just the degrees of freedom that correspond to the specified target deformations on the boundary $\Gamma_T$. Hence, the diagonal entries of $J$ are simply the indicator function for $\Gamma_T$. The constraint ensures that the deformed state is in static equilibrium.

**Simultaneous Topology and Heating Pattern Optimization**
Optimizing only the material densities $\rho$ is not sufficient to enable multi-state deformation resulting from a single input actuation. Rather the optimization must include the possibility to take advantage of the temperature varying stiffness of the SMP material. This is accomplished by introducing an additional optimization variable, $\eta : \Omega \to [0,1]$, which encodes the thermal state of the material at each point in the object. Here, $\eta = 0$ represents the material at room temperature, while $\eta = 1$ indicates that the material is heated above the $T_g$. In combination with the material density, $\rho$, which controls the presence or absence of material (12), the following material interpolation scheme is used for computing the Young's modulus:

$$E(\rho, \eta) = \int_\Omega \rho[(1-\eta)E_{\min} + \eta E_{\max}] \tag{4}$$

where $E_{\min}$ and $E_{\max}$ are the minimum and maximum stiffness of the material, which correspond to temperatures above and below the $T_g$, respectively. This yields a combined topology and material optimization formulation, where every element has the two design variables $\rho$ and $\eta$. To prevent the stiffness matrix to become singular, the minimum density value is set to $\rho = 0.001$ as commonly described in literature (12).

**Connected Topologies**
To ensure that the optimization algorithm generates structurally sound and connected topologies, a compliance term $C(\rho, \eta)$ is introduced to the objective function. This approach is similar to others found in literature (11, 13, 14) but adapted to the particular displacement-driven use case in this work. This additional term is required as many configurations exist that are kinematically valid and produce the correct target deformations, but are statically not necessarily feasible, as they for example include very thin connections or unconnected regions. To bias the optimization towards solutions that fulfill both criteria, the observation that connected, structurally sound topologies are in general stiffer than unconnected ones can be leveraged. Hence, additional compliance terms are introduced, one for each target deformation state, that measure the compliance in response to unit reaction forces generated by $u_T$. These reaction forces $f_C$ act in the opposite direction of $u_T$ and have normal magnitude. These additional terms maximize the stiffness between the target boundary $\Gamma_T = \Gamma_C$ and the fixed boundary $\Gamma_f$, ensuring that both regions are connected, and no unconnected regions are evolving. To comply with the overall minimization problem formulation, a minimum compliance formulation is used, which is equivalent to maximizing the stiffness. The resulting compliance energy $C$ is obtained by solving another linear elasticity problem for each target deformation state to compute $u_c$:

$$C(u) = \int_{\Gamma_c} f_c u_c \tag{5}$$
$$\text{with } K_{\rho,\eta} u_c = f_c$$

**Binary Structures**

Finally, it must be ensured that the output of the algorithm is a binary set of material/empty space assignment, i.e. $\rho \in \{0,1\}$. This is achieved by three modifications of the original optimization problem. First, the SIMP penalization is used with the parameter $p = 3$ to penalize intermediate densities and thermal states (15). This amounts to a simple modification of Eq. 4 to

$$E(\rho, \eta) = \int_\Omega \rho^p [(1 - \eta^p) E_{\min} + \eta^p E_{\max}] \tag{6}$$

However, this adaptation is not enough to promote binary designs, as it generally promotes "efficient" material usage for compliance-based optimization problems. As minimizing a compliance term is only a part of the problem formulation here, additional adjustments are necessary to promote fully solid-void-designs. Hence, a regularization term $R(\xi)$ is additionally introduced that penalizes the intermediate design variables $\xi \in\, ]\xi_{\min}, \xi_{\max}[$. The regularization takes the form of a quadratic function:

$$R(\xi) = -\frac{R_{\max}}{\left(\frac{(\xi_{\max} - \xi_{\min})}{2}\right)^2} \left(\xi - \frac{\xi_{\max} + \xi_{\min}}{2}\right)^2 + R_{\max} \tag{7}$$

Since the regularization term is 0 only at the bounds $\xi_{\min}$ and $\xi_{\max}$, a penalty is added to all intermediate values. If any intermediate values remain after the regularization, a projection $P(\xi)$ is applied. For some scalar value $\xi$, the projection maps all intermediate values $\xi \in ]\xi_{min}, \xi_{max}[$ to the space of binary values $\{\xi_{min}, \xi_{max}\}$

$$P(\xi) = \begin{cases} \xi_{min} & if \ \xi \leq \xi_t \\ \xi_{max} & otherwise \end{cases} \tag{8}$$

Here, the threshold $\xi_t \in (\xi_{min}, \xi_{max})$ is chosen such that the target deformation objective $J$ is minimized. This yields the optimization problem formulation for $k$ target deformation states:

$$\min_{\boldsymbol{\rho},\boldsymbol{\eta}} \sum_{j=1}^{k} J_j(\boldsymbol{\rho}, \boldsymbol{\eta}_j) + \alpha \sum_{j=1}^{k} C_j(\boldsymbol{\rho}, \boldsymbol{\eta}_j) + \sum_{j=1}^{k} \sum_{i=1}^{n} R_i(\boldsymbol{\rho}, \boldsymbol{\eta}_j) \tag{9}$$
$$\text{s.t.} \quad \mathbf{K}_{\rho,\eta_j} \mathbf{u}_j = \mathbf{f}_j, \quad \forall j = 1 \dots k$$
$$\mathbf{K}_{\rho,\eta_j} \mathbf{u}_{c,j} = \mathbf{f}_{c,j}, \quad \forall j = 1 \dots k$$

where $\alpha$ is a weighting parameter as commonly used in multi-objective optimization problems, which is set to $\alpha = 1$ in this work.

## Adaptive Finite Element Discretization

This section describes the Finite Element discretization, which lies at the heart of the proposed method. Although the method relies on standard linear elasticity, and displacement and traction boundary conditions are applied in the normal way, the geometric discretization is unique in that it is based on adaptive, volume constrained power diagrams. Typical approaches to topology optimization rely on extremely refined discretizations to capture geometric details, which can lead to a variety of artifacts (12). This specifically includes checkerboarding, hinging, and very small features that negatively influence the fabricability (16). To avoid these complications and to facilitate the actuation of the structures in this work, this approach relies on constructing a comparably small number of polygonal elements that can adapt themselves to find an optimized arrangement. Every polygon represents one individual cell with a dedicated material density and individual heating scheme. Hence, the size of the polygons is directly related to the size of the structural features within the design domain. The advantages of the polygon-based approach are highlighted in Fig. 2. The structure on the left is discretized by individual triangular elements, where every finite element has its own density and heating pattern. On the right, a structure partitioned by polygonal cells is shown, where the density and heating pattern is defined for every cell. While both structures achieve the same deformation behavior, the polygonal meshing reduces the overall number of variables, explicitly avoids high-resolution artifacts such as the checkerboarding, and hence improves the fabricability of the structure. Finally, making the geometry of the cells adaptive and including them as design variables in the optimization routine allows the underlying mesh to change its geometry and tune the shape of the resulting structures independent of the density distribution.

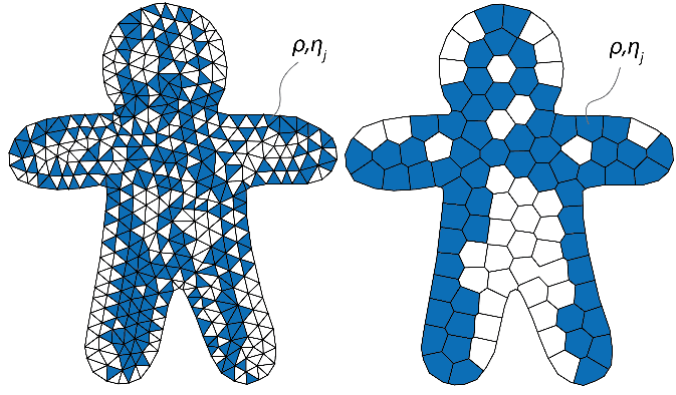

Fig. 2: Two optimized structures with solid-void density distribution. Left: Unstructured triangular mesh. Right: Adaptive power diagram cells

**Power Diagrams**

In this work, power cells are used to generate the adaptive polygonal partitioning of the design domain Ω. Power cells are a generalized version of Voronoi cells. Voronoi cells in 2D describe convex polygons that are based on a set of points in the Euclidean plane. In contrast to Voronoi cells, power cells can change their geometry and size by adding specific weights to the cells. Figure 3 shows a set of power cells and the corresponding nomenclature. A power diagram (9) partitions a domain Ω into cells, where every power cell $\mathcal{V}_i$ is defined by the respective power diagram site $x_i$ and weight $w_i$ via

$$\mathcal{V}_i^w = \left\{ x \in \Omega \mid \|x - x_i\|^2 - w_i \leq \|x - x_j\|^2 - w_j, \forall j \right\} \tag{10}$$

The two cells $\mathcal{V}_i$ and $\mathcal{V}_j$ are neighbors if the intersection $\mathcal{V}_i \cap \mathcal{V}_j$ is not empty (an edge in 2D or a planar polygon in 3D). The boundary between the two cells is denoted as $e_{ij}^*$. The Euclidean distance between $x_i$ and $x_j$ is $|e_{ij}|$ and the area of a power cell is $V_i$. Note that a power diagram with equal weights of all cells equals the Voronoi diagram with the same sites. The well known dualism between Voronoi and Delaunay cells also holds for power diagrams, where the dual cells constructed from connections between power sites orthogonal to the cell boundaries are called weighted Delaunay cells (17).

To increase the generality of the proposed method and include the design of structures with non-convex boundary shapes, the construction of power diagrams on non-convex domains must be adapted as power cells by definition are convex. For a given non-convex domain Ω, this is done by first constructing the normal power diagram with $n$ randomly initiated sites inside of the domain. Next, all cells that intersect with Ω are cut by applying the Sutherland-Hodgeman algorithm (18). For non-convex domains, this can result in the formation of non-convex power cells, from which all relevant parameters such as $V_i$, $e_{ij}^*$, and $e_{ij}$ can then be computed.

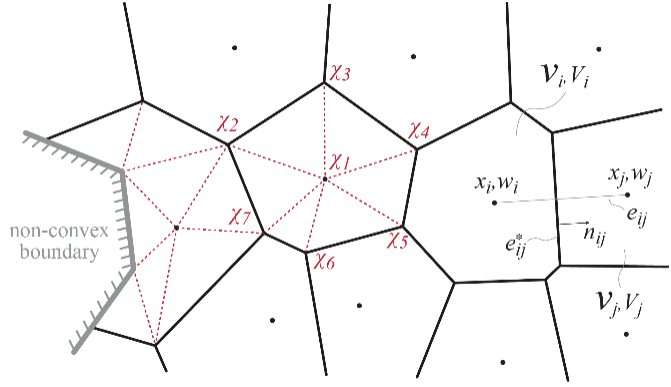

Fig. 3: Power diagram nomenclature and associated FE-mesh. The solid black lines show the boundaries of the power cells, the dotted red lines illustrate the FE-mesh, and the solid grey lines indicate the design domain boundary $\partial W$.

For structural computations, a triangular finite element mesh is then constructed from the clipped cells by connecting the two ends of every edge of a power cell $\mathcal{V}_i$ with the power cell site $x_i$. The cell boundary vertices and the cell sites form the vertices $\chi$ of the finite element mesh. For non-convex cells, not all boundary vertices can be connected to the cell sites. To build a triangular mesh in these cells without introducing additional points and edges, a classic Delauney triangulation (19) is employed that only uses cut-polygon edges and the cell site. The dotted red lines in Figure 3 show the triangular mesh constructed from convex and clipped non-convex power diagram cells. In the optimization framework, the same material density and heating scheme is assigned to all triangular finite elements in one cell. Hence, the number of cells directly links to the number of design variables.

This meshing approach further simplifies the calculation of derivatives compared to other meshing strategies since the finite element mesh vertices are now implicitly linked to the construction of the (non-)convex power cells.

**Volume Constraints**

A key advantage of power diagrams over Voronoi diagrams is that the cell sizes can be changed without having to move the cell sites. This is useful as by constraining the size of the individual cells, a minimum feature size can be enforced. This is implemented by adjusting the weights $w_i$ of the power cells. It was shown by (20) that finding a set of weights that enforces the cell volumes $V_i$ to match a given set of target cell volumes $V_{t,i}$ is equivalent to minimizing the smooth energy

$$\mathcal{E}(X,W) = \sum_i \int_{\mathcal{V}_i} \|x - x_i\|^2 dx - \sum_i w_i(V_i - V_{t,i}) \tag{11}$$

De Goes et al. (21) show that minimizing this energy is equivalent to solving an optimal transport problem, a well-studied problem in computer graphics, with volume constraints. The set of weights $W$ is found by updating the power diagram via Newton iterations and the derivatives have simple, closed-form expressions. It is further shown that this directly relates to finding a centroidal power diagram. Centroidal power diagrams are power diagram where each site $x_i$ coincides with the centroid, i.e. the center of mass, of each cell.

Since a volume-constrained power diagram is unique for a given set of target volumes $V_{t,i}$, the geometry of the power cells can be inversely controlled by prescribing different target volumes. This behavior can be used in the overall optimization framework proposed here by treating the target volumes as design variables to allow the power diagram, and hence the finite element mesh, to adapt during the optimization. Limiting the minimum and maximum cell volume to $V_{\min} \leq V_i \leq V_{\max}$ gives direct control over the feature size of the resulting geometry, enabling smooth, fabrication-ready structures. The algorithm to produce centroidal, volume-constrained power diagrams based on (21) is briefly summarized in Algorithm 1. The threshold for the gradient norm, which directly measures how close the sites are to the actual centroid of the polygons, is computed via $\epsilon_{VCPD} = 10^{-4} L_{\text{domain}} V_{\text{mean}} \sqrt{8n}$. This ensures that the mean error between the cell sites and the cell centroids is related to the overall size of the problem and is on average not larger than $10^{-4} L_{\text{domain}}$, where $L_{\text{domain}}$ is the maximum length of the domain $\Omega$. To ensure that the individual cell volumes $V_i$ always add up to the total domain volume $V_{\text{total}}$, the relative cell volume $\phi$ is introduced. For a given set of variables $\phi_i$, the physical cell volumes $V_i$ can be computed via

$$V_i = \frac{\phi_i}{\sum_{i=1}^{n} \phi_i} V_{\text{total}} \qquad (12)$$

---

**Algorithm 1:** Centroidal volume-constrained power diagram

**Data:** domain $\Omega$, target volumes $V_t$, number of cells $n$, initial positions of sites $X$

**Result:** Centroidal volume-constrained power diagram

1 **while** $\|\nabla_X \varepsilon\| \geq \epsilon_{\text{VCPD}}$ **do**
2     Enforce volume constraint;
3     Lloyd-step or gradient-descent step;
4     Update power diagram;
5 **end**

---

## Final Optimization Problem

The power diagrams are integrated into the topology optimization by explicitly updating the power diagram at the beginning of every optimization step according to the relative cell volume design variables $\phi$. After that, the finite element mesh is constructed and the objective function is computed. The optimization problem described in Eq. 9 is modified to become

$$\min_{\boldsymbol{\rho},\boldsymbol{\eta},\boldsymbol{\phi}} \mathcal{F} = \sum_{j=1}^{k} J_j(\boldsymbol{\rho},\boldsymbol{\eta}_j,\boldsymbol{\phi}) + \alpha \sum_{j=1}^{k} C_j(\boldsymbol{\rho},\boldsymbol{\eta}_j,\boldsymbol{\phi}) + \sum_{j=1}^{k}\sum_{i=1}^{n} R_i(\boldsymbol{\rho},\boldsymbol{\eta}_j) \quad (13)$$

$$\text{s.t.} \quad \mathbf{K}_{\rho,\eta_j,\phi}\mathbf{u}_j = \mathbf{f}_j, \quad \forall j = 1\ldots k$$

$$\mathbf{K}_{\rho,\eta_j,\phi}\mathbf{u}_{c,j} = \mathbf{f}_{c,j}, \quad \forall j = 1\ldots k$$

$$\sum_{i=1}^{n} V_i(\boldsymbol{\phi}) = V_{\text{total}}$$

$$\text{with} \quad J_j = \|\mathbf{L}_{u,j}(u_{\text{T},j} - u_j)\|, \quad \forall j = 1\ldots k$$

$$C_j = f_{c,j}^T u_{c,j}, \quad \forall j = 1\ldots k$$

$$R_i(\boldsymbol{\rho},\boldsymbol{\eta}_j) = R_i(\boldsymbol{\rho}) + R_i(\eta_j), \quad \forall j = 1\ldots k, \forall i = 1\ldots n$$

To minimize this objective, the gradients of the power diagram mesh with respect to the relative power cell volumes $\boldsymbol{\phi}$ are required, which are provided in the Supplementary Information. The full optimization algorithm is summarized in Algorithm 2.

---

**Algorithm 2:** Multi-state topology optimization

**Data:** domain $\Omega$, boundary conditions, $k$ target deformations $u_{\text{T},j}$, number of power diagram cells $n$

**Result:** Binary material density distribution; heating scheme to achieve different target deformations with a single structure

1. Initialize power diagram sites $X$ with $n$ random points inside the domain $\Omega$;
2. Initialize normalized cell sizes with $\phi_i = 1$;
3. Initialize cell densities with $\rho_i = 0.5$;
4. Initialize temperature with $\eta_i = 0.5$;
5. Initialize $R_{\max} = 0$;
6. **while** $\|\nabla_{\rho,\eta,\phi}\mathcal{F}\|_\infty \geq 10^{-6}$ and $n_{\text{iter}} \leq n_{\text{iter,max}}$ **do**
7.     Compute volume-constrained power diagram;
8.     Compute derivatives;
9.     Update design variables;
10. **end**
11. Increase Rmax
12. **while** $\|\nabla_{\rho,\eta,\phi}\mathcal{F}\|_\infty \geq 10^{-6}$ and $n_{\text{iter}} \leq n_{\text{iter,max}}$ **do**
13.     Compute derivatives;
14.     Update design variables;
15. **end**
16. Apply projection scheme to obtain binary values;

---

The two-step approach allows the algorithm to converge to a general solution first without applying the regularization. Such continuation schemes where intermediate densities are not fully penalized right from the beginning are common in topology optimization (12). Hence, the first step in the optimization is used to find a find a structural configuration of power cells along with a material distribution that minimizes the first two terms of the objective function. In a second step, the regularization term is added, but the power diagram is fixed. This step penalizes intermediate material densities and intermediate temperatures and promotes discrete solutions. MATLAB's *fmincon* interior point algorithm with an L-BFGS Hessian approximation is used in both steps to solve the optimization problem. If any non-discrete material densities or temperature values remain, the projection scheme is applied.

## Material Dithering

To fabricate the multi-state structures resulting from the optimization framework, a Stratasys Objet500 Connex3 3D printer (Stratasys, Eden Prairie, MN, USA) is used. Multi-state fabrication can be achieved by exploiting the differences in Young's modulus $E$ and glass transition temperature $T_g$ between the most rigid material VeroWhite+ (VW) and the most compliant material Agilus30 (AG). VW has a glass transition temperature of about $T_g \approx 65\ °C$ and hence is rigid at room temperature ($23\ °C$) with $E \approx 2100\ MPa$. Upon heating above its $T_g$ to $70\ °C$, the Young's modulus is reduced to about $8\ MPa$. AG has a glass transition temperature of $T_g \approx 15\ °C$ and is at room temperature already in its rubbery state with $E \approx 0.8\ MPa$. Upon heating it to $70\ °C$, the Young's modulus is further reduced to about $0.2\ MPa$. The multi-state approach requires a material that has a large change in stiffness at elevated temperatures but is compliant enough so that it can be deformed by hand at room temperature. As this is not directly achieved by any of the two base materials or any of the off-the-shelf intermediate, digital materials provided by the 3D printer, a manually programmed voxel-based dithering algorithm is used in this work. By dithering the two base materials at a voxel level (22, 23), the mechanical and thermal properties of the resulting material can be tuned.

To produce voxelized materials, the solid regions of the printable objects are divided into cuboids with edge lengths of $0.25 \times 0.25 \times 0.40\ mm^3$. Each cuboid represents one voxel and is randomly assigned one of the two base materials (VW or AG), where the overall percentage of both materials can be prescribed. The material properties of the resulting dithered materials are characterized in tensile tests according to the ASTM norm D638-14 (24). Figure 4A and B show the stress-strain curves of different dithered materials with 10%, 30%, 50%, 70%, and 90% AG at room temperature (A) and at $70\ °C$ (B), respectively. With increasing percentage of AG, the dithered material becomes significantly more compliant in both states. Figure 4C shows the Young's modulus at both temperatures for the different dithering ratios. Since the Young's modulus at high temperatures (red dotted line) is comparably low and does not change significantly over the temperature range as both materials are in their rubbery state, the absolute stiffness reduction by heating of a dithered material can be directly controlled by changing the mixture ratio. For the multi-state structures in this work, a mixture ratio of 50% is chosen, where the Young's modulus at room temperature of $E_{\text{RT}} \approx 120\ MPa$ drops to $E_{70} \approx 2.9\ MPa$ at $70\ °C$.

Once printed, loops of coated copper wire with a diameter of $0.15\ mm$ are used to differentially heat the structures. For every target deformation state, one continuous wire is used. The wire loops are fixed with super glue to the structure and are heated to about $70\ °C$ by applying a current of about $0.9\ A$. For actuation, the structures are fixed to a base plate with metal nails and out of plane deformations are prevented by an acrylic plate on top of the structures. The input actuation displacement is manually applied with an acrylic bar.

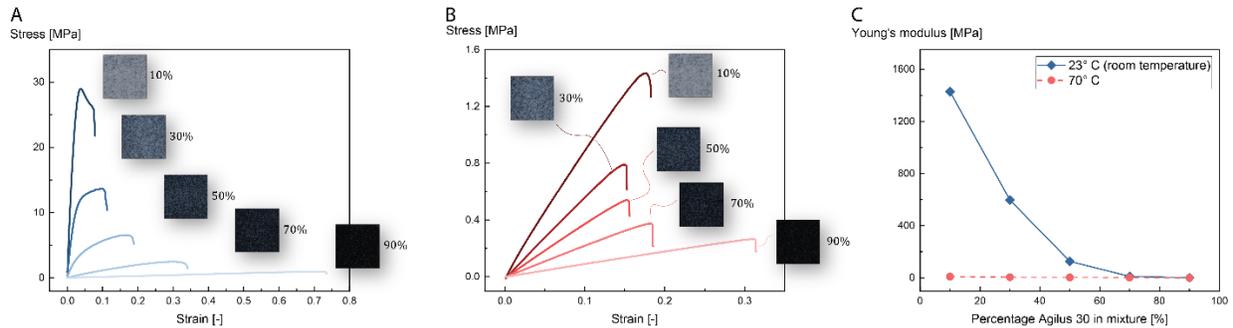

Fig. 4: Material data for dithered materials. (A) Stress-strain-curves at room temperature (T = 23 °C) of dithered materials for different percentages of Agilus 30. (B) Stress-strain-curves at increased temperature (T = 70 °C) of dithered materials for different percentages of Agilus 30. The insets show images of 3D-printed material samples. (C) Young's modulus at room temperature and at 70 °C for the different material compositions.

## RESULTS

The previously described optimization framework is now used to generate different example structures that show its potential for applications in the field of computer graphics and mechanical engineering. Figure 5 shows the previously introduced "Gingerbread Man". On the left, the optimization problem setup is depicted. The green hashes show fixed boundary conditions, the blue arrows indicate the position and direction of the input actuation displacement. The orange arrows indicate two different sets of target displacements, which are prescribed at the points marked with "a" and "b". The "Gingerbread Man" tilts to left or to the right when its head is pushed downwards for target state 1 and target state 2, respectively. In the middle of Fig. 5, the optimized structures are shown in the initial state, both deformed states, as well as overlaid with the "Gingerbread Man" visualization. Blue cells indicate the presence of solid material and the cells that are heated in each state are highlighted by an orange hatching. The solid green lines show the original shape and position of the target regions, i.e. the "Gingerbread Man's" arms here, for reference. The deformed "Gingerbread Man" visualization is obtained by applying the displacement field computed by the FE-simulations to the original image of the undeformed "Gingerbread Man". On the right side of Fig. 5, the fabricated "Gingerbread Man" structure is shown in the initial state and in both deformed target states. The heated regions in both states are highlighted in red and the initial positions of the arms are indicated by the solid green lines, respectively.

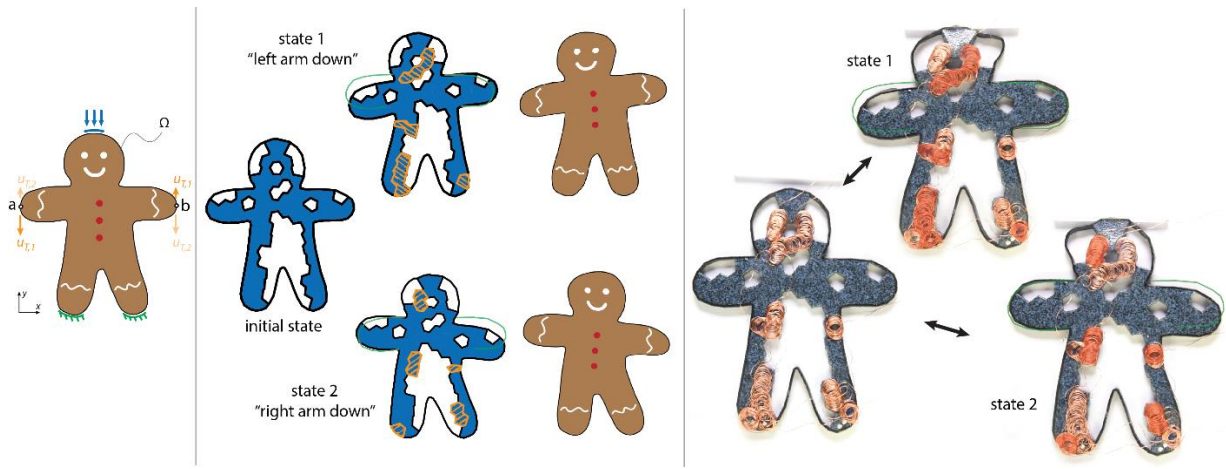

Figure 5: A multi-state "Gingerbread Man" that move both its left arm and its right arm downwards in response to an identical actuation. Left: Design domain where green hashes show fixed boundary conditions, blue arrows indicate actuation position and direction, and $u_{T,1}$ and $u_{T,2}$ indicate the target displacements for the first and second states. Center: Optimized material distribution in the initial and deformed states with corresponding local heating regions shown by hashed orange cells. The green contour shows the reference configuration. Right: Fabricated result deformed into both states.

Here it becomes clear how a single, repeated actuation can produce two diametrically opposed motions by sequential heating of the respective regions. In both these cases, control of the motion is direction encoded into the material rather than the actuator, which significantly simplifies actuator control, as only a linear input displacement is required. While the tilting motion in both directions is qualitatively well captured by both the simulated, optimized structures and the experiments, the magnitude of the deformation is slightly larger in the simulations. To quantitatively compare the accuracy of both optimization and experiments, the values of the prescribed target displacements ($u_T$) and of the displacements achieved in simulations ($u_{\text{sim}}$) and experiments ($u_{\text{exp}}$) are provided in Tab. 1 for the "Gingerbread Man" and all subsequent examples.

The experimental deformations are measured via the image processing software ImageJ (25). For brevity, the table provides only the magnitudes of the displacements. The direction of the displacements is indicated by the orange arrows in Fig. 5 and 6, respectively. For completeness, the values of the driving input actuations ($u_{\text{in}}$) are further provided. For the "Gingerbread Man" example, it can be seen that the target displacement of $8\ mm$ at the points "a" and "b" is matched by the simulation with deviations of maximum $0.4\ mm$ in both states, whereas the experimentally tested structures only achieve a downward deflection of $5.0\ mm$ and $6.8\ mm$, respectively.

Table 1: Values of the input displacements $u_{in}$, target deformations $u_{T,j}$ of all target states, and deformations of the target domain from simulations ($u_{sim}$) and experiments ($u_{exp}$). The positions a, b, c and the direction of the deformations are shown in the Figures 5 and 6, respectively. All values are in mm.

|  | Gingerbread | | Airfoil | Armadillo | | Dinosaur | | |
| --- | --- | --- | --- | --- | --- | --- | --- | --- |
|  | a | b | Tip | a | b | a | b | c |
| $u_{in}$ | 5.0 | | 5.0 | 7.0 | | 4.0 | | |
| $u_{T,1}$ | 8.0 | 2.0 | 11.0 | 0.0 | 3.0 | 5.0 | - | - |
| $u_{T,2}$ | 2.0 | 8.0 | 11.0 | 3.0 | 0.0 | 5.0 | - | - |
| $u_{T,3}$ | - | - | - | - | - | - | 4.0 | 4.0 |
| $u_{sim,1}$ | 8.2 | 2.2 | 12.7 | 0.0 | 1.9 | 3.5 | - | - |
| $u_{sim,2}$ | 2.1 | 7.6 | 10.8 | 1.7 | 0.0 | 4.9 | - | - |
| $u_{sim,3}$ | - | - | - | - | - | - | 2.8 | 3.1 |
| $u_{exp,1}$ | 5.0 | 2.02 | 10.3 | 0.0 | 1.8 | 2.9 | - | - |
| $u_{exp,2}$ | 2.0 | 6.8 | 9.1 | 1.7 | 0.1 | 4.8 | - | - |
| $u_{exp,3}$ | - | - | - | - | - | - | 2.3 | 2.0 |

Figure 6 shows three more examples to illustrate the optimization approach. In Fig. 6A, the back part of a shape-morphing "Airfoil" is shown. When the input actuation is applied, the tip of the "Airfoil" can both deflect up and down based on the pattern of localized heating. Table 1 shows that both simulated and experimentally tested structures match the target deformations with maximum deviations of $1.7\ mm$ and $1.9\ mm$, respectively. Figure 6B shows a classic example from the field of computer graphics. The "Armadillo" can lower its right arm or its left arm when its right leg is moved in the direction of the blue arrow. While the previous examples featured wide, round boundaries, this example is characterized by small features and a "narrow" overall design space. Nonetheless, the power cells can form connected structures that qualitatively show the intended behavior. However, the absolute values of the output displacements for both the simulated structure and the fabricated structure reach only about 2/3 of the magnitude of the prescribed target displacement, as seen in Tab. 1. Finally, Fig. 6C shows a structure that can reach three individual states with a single input actuation. The "Dinosaur" can move its tail up and down and close its mouth when it is pushed on its back. Table 1 shows that while the prescribed target downward movement of the tail of $5.0\ mm$ is almost fully reached in simulation and experiment, the discrepancy between target motion and simulation and experiment is much larger for the upward motion of the tail and for the mouth-closing motion. Nonetheless, the basic motions are captured by the optimized structure and show that it is possible to encode up to three different deformation patterns in a structure with a single actuation input.

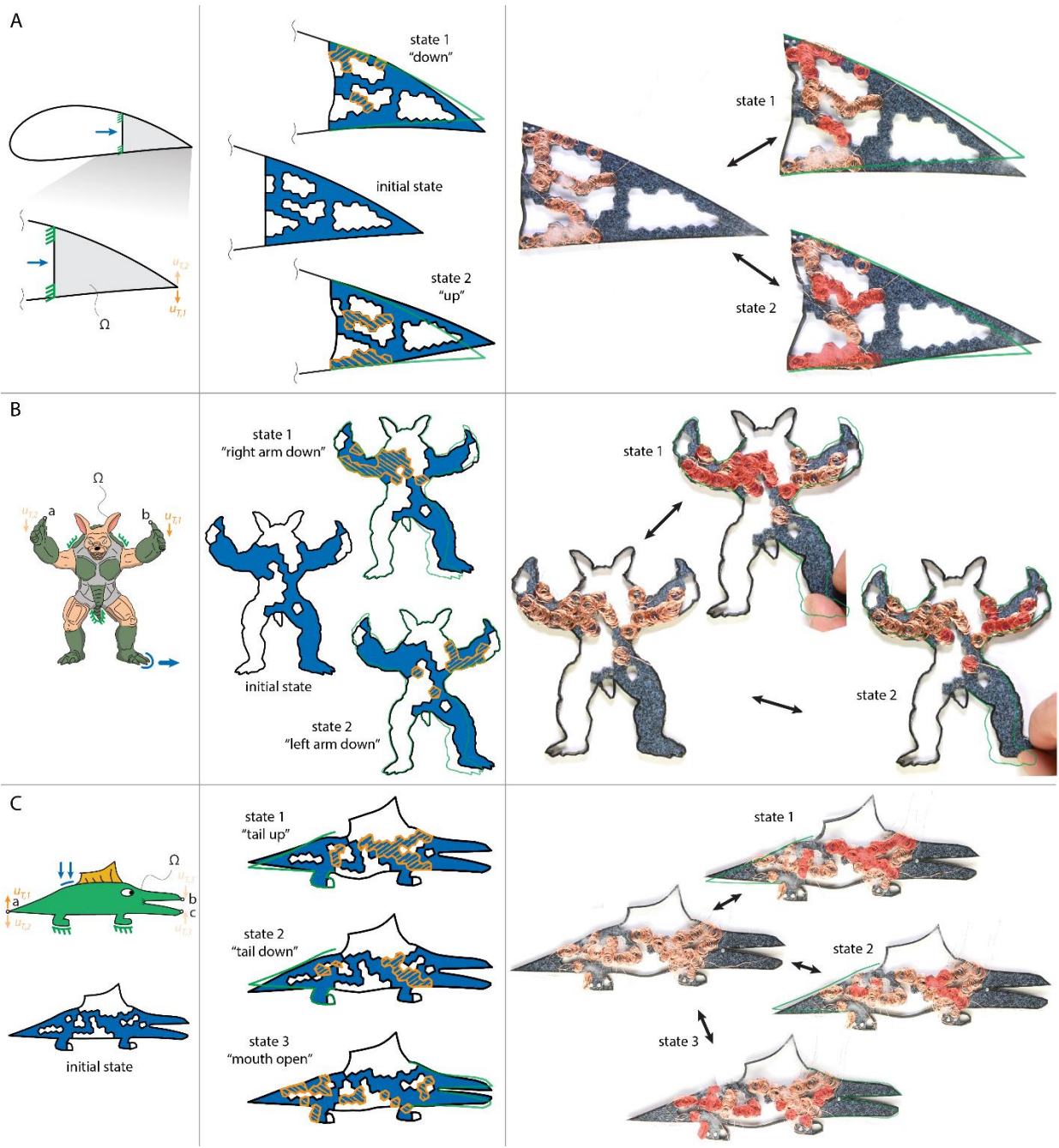

Figure 6: Additional results produced by the multi-state design and fabrication process. (A) "Airfoil" that can move its tip up an down. (B) "Armadillo" that can move the right arm and the left arm down.  C) "Dinosaur" with three target deformations. It can move its tail up and down and open its mouth from a single actuation on its back.

## DISCUSSION

The results show that the proposed algorithm can produce structures that can achieve multiple different target deformations with only a single input actuation displacement. The mechanisms and heating patterns that emerge from the optimization feature linkage type bars and joints that drive the deformation. These features are all mechanically viable and no unconnected regions appear. The heating patterns act as switches, allowing force to flow through various paths in these linkage graphs. This can be observed for example in the "Airfoil" example in Fig. 6A, where the activated heating in state 1 "disconnects" the upper fixed region of the "Airfoil", allowing it to move downwards. The activated heating in state 2 "disconnects" the lower fixed part of the "Airfoil" and it deflects upwards. Similar observations can be made for the other examples, where activated heating locally softens the material and enables deformations.

While all structures qualitatively capture the encoded motions, the achieved magnitude of the deformations is generally small and differs between the examples. The overall small magnitude of the deformations stems from the linear-elastic finite element approach as well as from the associated materials themselves. By definition, the deformations that can be accurately captured by a linear-elastic numerical model are small and follow linear deformation paths. To capture geometrically more complex deformations or allow larger, non-linear strains, different material models would have to be considered. This would also involve more complex derivatives and further complicate the optimization problem. Additionally, the dithered material used in this work has a very low failure strain of $< 5\%$ at low temperatures. As the structures in this work rely on the tailored properties of the dithered material, which are currently not found in any other commonly available 3D printing material, this further limits the overall achievable deformations of the structures. However, this limitation could be overcome when a wider range of 3D printable materials with different or tunable thermo-mechanical properties will be available in the future. For these reasons, this work is currently restricted to linear-elastic, small deformations.

Besides the overall small magnitude of the deformations, the quantitative comparison in Tab. 1 reveals that the specified target deformations are not achieved equally well by all structures. The simulated and experimentally observed displacements of the optimized "Airfoil" structures reach the values of the prescribed target displacements with mean errors of about 10%, respectively. However, the quantitative analysis of the deformations shows that for the examples "Armadillo", "Smiley", and "Dinosaur", some of the simulated and experimentally observed displacements reach only about half of the specified target displacements. As this involves the accuracy of the optimization and the fabrication framework, both are discussed in the following paragraphs.

The computational framework combines the geometric mesh control via power diagrams with the material and heating pattern optimization to achieve the prescribed target deformations. The material and heating pattern optimization is by definition a discrete problem, where a specific material (solid or void) and a discrete temperature for each state (above $T_g$ or below $T_g$) are assigned to every cell. To solve this problem, a standard approach from topology optimization is employed where the relaxed, continuous problem is solved and penalization techniques are applied to bias the optimization towards discrete values. This allows for the use of gradient-based optimization algorithms. Figure 7A shows the effect of the regularization and projection step in this work.

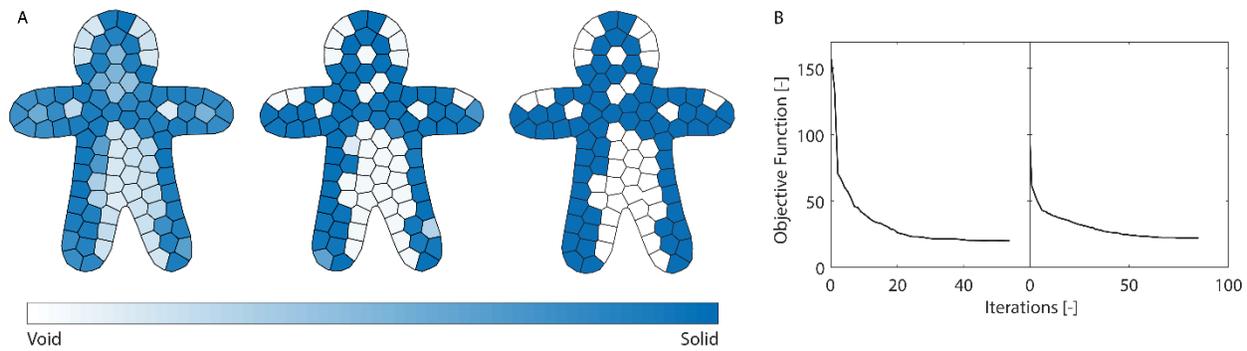

Figure 7: (A) Continuous densities, without regularization (left), almost discrete densities after regularization step (middle), and discrete densities after projection (right). (B) Convergence plot of the "Gingerbread Man" example.

The image on the left shows the optimized structure after the first optimization step. No regularization is applied yet, but intermediate densities are penalized by the SIMP parameter $p = 3$. As this parameter only affects the compliance terms $C_j$ of the objective function, intermediate densities remain. However, structural features emerge and the power diagram discretization ensures a minimum member size. In the second step, the regularization term $R$ is added to the objective function. The image in the middle shows that intermediate densities almost completely disappear. The final topology in the right image is obtained by applying the projection step and yields a binary material distribution.

The convergence of the first and second optimization step is shown in Fig. 7B. Both steps continuously converge towards a solution where the first-order optimality criterion is fulfilled. However, convergence to a global optimum can not be guaranteed with a local, gradient-based optimization method such as the interior point method in MATLAB's constrained optimization *fmincon*. While this is generally not a problem and the target deformations can be matched for simple examples such as the "Gingerbread Man" and the "Airfoil", the optimization algorithm has problems to achieve the specified target deformation for more complex problem setups where for example the design domain is narrow and has small features ("Armadillo") or more than two target states are specified ("Dinosaur"). This behavior could be further improved in future works by admitting different starting points or combining the local optimization for example with a stochastic search method such as pattern search. Nonetheless, the basic encoded deformations are visible for the local solutions and convey the intended characteristics.

The inaccuracies observed between the simulated structure and the experimentally deformed structure mostly stem from how the structures are heated in the experiments. The numerical model assumes that the temperature of each cell is either exactly room temperature or $70\ °C$. In reality, there are inaccuracies in the placement of the copper heating coils, which sometimes slightly extend to adjacent cells, and there is heat conduction within the material between adjacent cells. Further, the heat generated by the copper coils can only partially escape from below the acrylic glass such that a precise temperature control is not feasible for this setup. Even though global heating approaches can provide more precise control over the absolute temperature, they are unsuitable to provide temperature gradients within a structure, which are essential for activating different states for a single mechanical input as shown in this work. While the experimental heating is assumed to be the biggest source of error here, other fabrication inaccuracies and inaccuracies due to how the boundary conditions are applied in the experiments can occur additionally. Future improvements could include improving the fabrication of the heating coils as well as minimizing the number of heated cells. In this work, the heating coils are manually coiled by wrapping copper wire around a cylindrical mandrel. This yields a single coil, which is then manually glued to the structure. Using a pre-coiled heating wire and applying the same number of coils to each cell could help to improve the regularity and consistency of the heating. By extending the objective function to minimize the overall use of heating elements, unnecessary placement of heating wire, as for example observed in the protruding cell in state 2 of the "Armadillo" example, could be prevented, which would further simplify the heating scheme.

The proposed method has some promising directions for future work. The next step would be to extend the method to 3D. While 2D animations are visually compelling and plane strain mechanisms can be useful in engineering, a 3D implementation of the method would further broaden its applicability. This does not require any fundamental changes, as both the power diagram method and the optimization framework can be directly extended to 3D. Furthermore, the nature of the used materials limits the applicability of the structure. The SMPs are viscoelastic, which implies that the time to reset an object after actuation to its initial state can be delayed and is determined by the materials' viscosity. Due to the generality of the method, it is not limited to SMPs only. Other materials that provide reversible control over their stiffness on heating such as non-SMP thermoplastics can be used to trigger different states. Finally, the heating and cooling of the materials takes time and must be considered during actuation. Cooling, in particular, can take up to several minutes, since it is passive. Active cooling or more precise heating could further improve the performance of the structures.

## CONCLUSIONS

In summary, this work presents a method to design multi-state structures that can be actuated by a single input displacement. The target states are encoded in the structure itself via both material distribution and temperature control. The different target states can be activated by partial differential heating of the structures. The method is based on a new topology optimization method, which simultaneously optimizes the material distribution and the heating pattern. By employing an adaptive power diagram meshing algorithm, common numerical artifacts such as checkerboarding and minimum member size can be directly controlled, and the fabricability of the structures is ensured. The validity of the proposed method is evaluated by different examples from computer graphics and engineering. In contrast to most other multi-state structures shown in literature, the method in this work does not require a large number of actuation mechanisms, but relies only on a single input actuation and one heating pattern for each state. This makes it suitable for future applications that require specific motions under simplified input actuation such as animated fabrication in computer graphics, active structures such as airfoils, and object manipulation for robotics applications.

## BIBLIOGRAPHY

1. J. Pérez, *et al.*, Design and fabrication of flexible rod meshes. *ACM Trans. Graph.* **34**, 138:1-138:12 (2015).

2. O. Weeger, Y. S. B. Kang, S.-K. Yeung, M. L. Dunn, Optimal Design and Manufacture of Active Rod Structures with Spatially Variable Materials. *3D Print. Addit. Manuf.* **3**, 204–215 (2016).

3. B. Zhu, M. Skouras, D. Chen, W. Matusik, Two-Scale Topology Optimization with Microstructures. *ACM Trans. Graph.* **36**, 16 (2017).

4. Y.-W. Lee, H. Ceylan, I. C. Yasa, U. Kilic, M. Sitti, 3D-Printed Multi-Stimuli-Responsive Mobile Micromachines. *ACS Appl. Mater. Interfaces* (2020) https:/doi.org/10.1021/acsami.0c18221.

5. X. Kuang, *et al.*, Advances in 4D Printing: Materials and Applications. *Adv. Funct. Mater.* **29**, 1–23 (2019).

6. S. Akbari, *et al.*, Enhanced multimaterial 4D printing with active hinges. *Smart Mater. Struct.* **27** (2018).

7. G. Sossou, *et al.*, Design for 4D printing: Modeling and computation of smart materials distributions. *Mater. Des.* **181** (2019).

8. T. S. Lumpe, K. Shea, Computational design of 3D-printed active lattice structures for reversible shape morphing. *J. Mater. Res.* (2021) https:/doi.org/10.1557/s43578-021-00225-2.

9. F. Aurenhammer, Power Diagrams: Properties, Algorithms and Applications. *SIAM J. Comput.* **16**, 78–96 (1987).

10. Q. Ge, H. J. Qi, M. L. Dunn, Active materials by four-dimension printing. *Appl. Phys. Lett.* **103**, 10–15 (2013).

11. S. R. Deepak, M. Dinesh, D. K. Sahu, G. K. Ananthasuresh, A comparative study of the formulations and benchmark problems for the topology optimization of compliant. *J. Mech. Robot.* **1**, 1–8 (2009).

12. M. P. Bendsøe, O. Sigmund, *Topology Optimization: Theory, Methods and Applications* (Springer, 2003).

13. M. Frecker, N. Kikuchi, S. Kota, Topology optimization of compliant mechanisms with multiple outputs. *Struct. Optim.* **17**, 269–278 (1999).

14. F. Stöckli, K. Shea, Topology Optimization of Rigid-Body Systems Considering Collision Avoidance. *J. Mech. Des.* (2020) https:/doi.org/10.1115/1.4046076.

15. M. P. Bendsøe, O. Sigmund, Material interpolation schemes in topology optimization. *Arch. Appl. Mech.* **69**, 635–654 (1999).

16. C. Talischi, G. H. Paulino, A. Pereira, I. F. M. Menezes, Polygonal finite elements for topology optimization: A unifying paradigm. *Int. J. Numer. Methods Eng.*, 671–698 (2010).

17. F. De Goes, P. Memari, P. Mullen, M. Desbrun, Weighted triangulations for geometry processing. *ACM Trans. Graph.* **33** (2014).

18. I. E. Sutherland, G. W. Hodgman, Reentrant Polygon Clipping. *Commun. ACM* **17**, 32–42 (1974).

19. L. P. Chew, Constrained delaunay triangulations. *Algorithmica* **4**, 97–108 (1989).

20. F. Aurenhammer, F. Hoffmann, B. Aronov, Minkowski-type theorems and least-squares clustering. *Algorithmica* **20**, 61–76 (1998).

21. F. De Goes, K. Breeden, V. Ostromoukhov, M. Desbrun, Blue Noise through Optimal Transport. *ACM Trans. Graph.* **31**, 1–12 (2012).

22. K. Vidimce, S. Wang, J. Ragan-Kelley, W. Matusik, OpenFab : A Programmable Pipeline for Multi-Material Fabrication. *ACM Trans. Graph.* **32**, 12 (2013).

23. C. Bader, D. Kolb, J. C. Weaver, N. Oxman, Data-Driven Material Modeling with Functional Advection for 3D Printing of Materially Heterogeneous Objects. *3D Print. Addit. Manuf.* **3**, 71–79 (2016).

24. ASTM International, "Standard Test Method for Tensile Properties of Plastics D638-14" (2004) https://doi.org/10.1520/D0638-14.1.

25. W. S. Rasband, ImageJ.